\documentclass[aps,prd,12pt,a4paper,groupedaddress,preprintnumbers,floatfix,nofootinbib,showpacs]{revtex4}
\usepackage[english]{babel}
\usepackage{amsmath}
\usepackage{graphicx}
\usepackage{color}
\usepackage{amssymb}
\usepackage{hyperref}
\usepackage{dcolumn}
\usepackage{bm}
\usepackage{longtable}

\usepackage{appendix}
\usepackage{float}

\begin{document}
\preprint{IFIC/14-71}
\title{NON-DIAGONAL CHARGED LEPTON MASS MATRIX,
 THE TBM AND NON-ZERO $\theta_{13}$}
\author{Alma D. Rojas\footnote{{Poster presenter. Co-authors: J.Alberto Acosta, Alfredo Aranda and Manuel A. Buen-Abad.}\\
\emph{Email address:} alma.rojas@ific.uv.es (Alma D. Rojas)\\
Contribution to the proceedings of the 37th International Conference on High Energy Physics (ICHEP 2014), 2-9 July 2014, Valencia, Spain.}}

\affiliation{AHEP Group, Instituto de F\'isica Corpuscular – C.S.I.C./Universitat de Val\`{e}ncia.\\
 C/ Catedr\'atico Jos\'e Beltr\'an, 2, E- 46980 Paterna (Val\`{e}ncia), Spain.}

\date{\today}

\begin{abstract}
Assuming that the neutrino mass matrix is diagonalized by the TBM, we looked for the charged lepton mass matrix textures which render a lepton mixing matrix consistent with data. We were interested in the textures with the maximum number of zeros, so, we explored the cases of real matrices with three (and also four zeros) and found which of them provide solutions in agreement with data. We present the successful Yukawa textures and obtained the relative sizes of their non-zero entries. We found some interesting relations among the entries of these textures in terms of the charged lepton masses. Complete details can be found in \cite{Acosta:2012qf}. 
\end{abstract}
\maketitle

\section{Introduction}
\label{Introduction}
In order to explain the flavor structure of Yukawa couplings in the Standard Model 
(SM), discrete flavor symmetries have been extensively used. 
The freedom to choose the Yukawa matrix structures has lead model builders to study 
some particular textures, for instance, the Nearest Neighbor Interaction (NNI) form (see \cite{NNI} 
 and references therein), 
the Fritzsch-like~\cite{Fritzsch} and non Fritzch-like textures (see \cite{Gupta:2011zzg} for an overview of both cases)
 and n-zero textures, among others. 

In the lepton sector, the matrix which contain the three mixing angles together with 
a CP violating phase is the lepton mixing matrix $U_{PMNS}$, defined as $U_{PMNS}=U_l^{\dagger}U_{\nu}$.
It can be parametrized in different ways, one of them is the standard parametrization~\cite{PDG}:
\small
\begin{equation}\label{upmnsPDG}
\left(\begin{array}{ccc}
 c_{12}c_{13}&s_{12}c_{13}&s_{13}e^{-i\delta}\\
 -s_{12}c_{23}-c_{12}s_{23}s_{13}e^{i\delta}&c_{12}c_{23}-s_{12}s_{23}s_{13}e^{i\delta}&s_{23}c_{13}\\
 s_{12}s_{23}-c_{12}c_{23}s_{13}e^{i\delta}&-c_{12}s_{23}-s_{12}c_{23}s_{13}e^{i\delta}&c_{23}c_{13}
\end{array}\right)
\end{equation}
\normalsize
where $s_{ij}=\sin\theta_{ij}$, $c_{ij}=\cos\theta_{ij}$ and $\delta$ is the Dirac CP violating phase.

The most used ansatz for the lepton mixing matrix  had been the tribimaximal mixing (TBM) 
matrix~\cite{Harrison:2002er}:
\small
\begin{equation}
U_{TBM} =\left(\begin{array}{ccc}
 \sqrt{\frac{2}{3}}&\frac{1}{\sqrt{3}}&0\\
 -\frac{1}{\sqrt{6}}&\frac{1}{\sqrt{3}}&-\frac{1}{\sqrt{2}}\\
-\frac{1}{\sqrt{6}}&\frac{1}{\sqrt{3}}&\frac{1}{\sqrt{2}}
\end{array}\right),
\end{equation}
\normalsize
which implies a zero value for $\theta_{13}$.
However, after the experimental confirmation of its non-zero value~\cite{nonzerot13exp}, the TBM requires corrections coming possibly from the charged lepton sector.

\section{Approach}

If we demand $U_{\nu}=U_{TBM}$ to be the matrix which diagonalizes $M_{\nu}$ and $U_{l}$ the one that  
diagonalizes $M_l^2= M_l M_l^{\dag}$ (in the weak basis), we wanted to determine the form of $M_l$ such that the $U_{PMNS}$ 
has values in the allowed experimental range, being
\small
\begin{equation}
 U_{PMNS}=U_l^{\dagger}U_{TBM}\quad \Rightarrow \quad U_{l}=U_{\nu}U_{PMNS}^{\dagger},
\end{equation}
\normalsize
and  $U_l$ is such that
\small
\begin{equation}\label{Msquare}
 M^{2}_{l}=U_{l}M^{2}_{lD}U_{l}^{\dagger},\qquad M^{2}_{lD} \equiv M_{lD}M^{\dagger}_{lD}
\end{equation}
\normalsize
where $M_{lD} = {\rm{diag}}(m_e,  \ m_{\mu},  \ m_{\tau})$, 
and $M_l$ is an arbitrary mass matrix with unknown values
\small
\begin{equation}\label{generalMl}
M_{l}=\left(\begin{array}{ccc}
 a&b&c\\
 d&e&f\\
g&h&i
\end{array}\right).
\end{equation}
\normalsize
In this first approach we considered the particular case of a real mass matrix, then $M_l^2$ takes the form
\small
\begin{equation}\label{realMl2}
 \left(\begin{array}{ccc}
 a^{2}+b^{2}+c^{2}&ad+be+cf&ag+bh+ci\\
 ad+be+cf&d^{2}+e^{2}+f^{2}&dg+eh+fi\\
ag+bh+ci&dg+eh+fi&g^{2}+h^{2}+i^{2}
\end{array}\right).
\end{equation}
\normalsize
We were interested in determining the textures with the maximum number of zeros 
(having such textures can be useful to model builders) because zeros may be due to some underlying flavor symmetry. 
So, we started looking for all the possible three-zero textures and we found sixteen different matrices (up to column permutations) shown in (\ref{texturas}).
\small
\begin{eqnarray}\label{texturas} 
&M_{301}=\left(\begin{array}{ccc} 
0 & 0 & c \\
d & e & 0 \\
g & h & i 
\end{array} \right); \
\quad 
M_{302}=\left(\begin{array}{ccc} 
0 & 0 & c \\
d & e & f \\
g & h & 0 
\end{array} \right);\
\quad
M_{303}=\left(\begin{array}{ccc} 
0 & b & c \\
d & 0 & 0 \\
g & h & i 
\end{array} \right);\
\checkmark M_{304}=\left(\begin{array}{ccc} 
0 & b & c \\
d & 0 & f \\
g & h & 0 
\end{array} \right);\nonumber\\
&M_{305}=\left(\begin{array}{ccc} 
0 & b & c \\
d & e & f \\
g & 0 & 0 
\end{array} \right); \
\quad
M_{306}=\left(\begin{array}{ccc} 
a & b & c \\
0 & 0 & f \\
g & h & 0 
\end{array} \right); \
\quad
M_{307}=\left(\begin{array}{ccc} 
a & b & c \\
0 & e & f \\
g & 0 & 0 
\end{array} \right); \
\checkmark M_{308}=\left(\begin{array}{ccc} 
0 & b & c \\
0 & e & f \\
g & 0 & i 
\end{array} \right); \nonumber\\
& \checkmark M_{309}=\left(\begin{array}{ccc} 
0 & b & c \\
0 & 0 & f \\
g & h & i 
\end{array} \right); \
\checkmark M_{310}=\left(\begin{array}{ccc} 
a & 0 & 0 \\
d & e & 0 \\
g & h & i 
\end{array} \right); \
 \checkmark M_{311}=\left(\begin{array}{ccc} 
0 & b & c \\
d & e & f \\
0 & 0 & i 
\end{array} \right); \
\checkmark M_{312}=\left(\begin{array}{ccc} 
0 & b & c \\
d & 0 & f \\
0 & h & i 
\end{array} \right); \nonumber\\
& \checkmark M_{313}=\left(\begin{array}{ccc} 
0 & 0 & c \\
d & e & f \\
0 & h & i 
\end{array} \right); \
\checkmark M_{314}=\left(\begin{array}{ccc} 
a & b & c \\
0 & e & f \\
0 & 0 & i 
\end{array} \right);\
 \checkmark M_{315}=\left(\begin{array}{ccc} 
a & b & c \\
0 & 0 & f \\
0 & h & i 
\end{array} \right); \
\checkmark M_{316}=\left(\begin{array}{ccc} 
a & 0 & c \\
0 & e & f \\
0 & h & i 
\end{array} \right).\nonumber\\ 
\end{eqnarray}
\normalsize
\section{Analysis and Results}
\begin{figure}[ht]
 \centering
 \includegraphics[width=6cm]{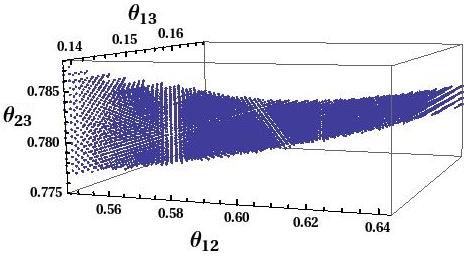}
 \caption{Solution volume of $M_{304}$ for $\delta=\pi$. 
 Similar results are obtained for $\delta=0$.}
 \label{SolVol}
\end{figure}

To perform the analysis, we made a scan over the allowed experimental range of the mixing angles taken from \cite{nonzerot13exp,Schwetz:2011zk}, 
 looking for the mixing angles combinations that provided real solutions to equation (\ref{Msquare}) (for the charged lepton masses we used the values in \cite{PDG} and set $\delta_{CP}$ values to $0$ and $\pi$). 
For textures $M_{301},M_{302},M_{303},M_{305},M_{306},M_{307}$ we did not find any solution. That leave us with ten possible textures.
For each texture of the Yukawa matrix, we find the maximum and minimum orders of 
magnitude of their entries. We also searched for the solution volume of each three-zero 
texture (i.e. the set of points given by the three mixing angles that make possible 
to find real solutions to the entries of the $M_l$ matrix). 
In nine of the ten cases the solution volume fills the complete experimentally allowed parameter space.
The most interesting case is that of texture $M_{304}$ (with zeroes on the diagonal), 
 which give us the solution volume showed in Figure \ref{SolVol}. 
The main characteristic of this case is that the angle $\theta_{23}$ is very restricted.
Its allowed interval is [0.7763,0.7876] for  $\delta_{CP}=\pi$, and [0.7750,0.7873] for $\delta_{CP}=0$. 
 These intervals exclude the best fit value of  $\theta_{23}$.

\begin{figure}[ht]
 \centering
 \includegraphics[height=6cm]{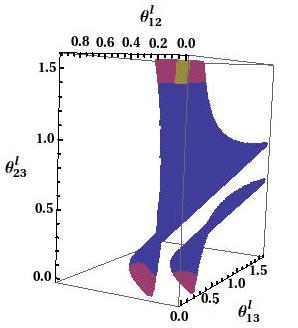}
 \caption{Sets of values for the angles of the charged lepton mixing matrix that match the experimental results. The yellow (light) band in the upper part of the plot is the region consistent with all experimental results. 
 (See \cite{Acosta:2012qf} for more details)}
 \label{MatchMixAngles}
\end{figure}

We found that there are extra conditions on the entries of  $Y_l=\frac{1}{\langle H \rangle} M_l$ (being $\langle H \rangle\approx 246$GeV the Higgs vev) in terms of the charged lepton masses:
\small
\begin{eqnarray}\label{YukawasRel}
 (Y_l Y_l^T)_{11}=y^{2}_{a}+y^{2}_{b}+y^{2}_{c}&=&7.7\times10^{-7} \nonumber\\
 (Y_l Y_l^T)_{22}=y^{2}_{d}+y^{2}_{e}+y^{2}_{f}&=&5.2\times10^{-5} \\
 (Y_l Y_l^T)_{33}=y^{2}_{g}+y^{2}_{h}+y^{2}_{i}&=&2.1\times10^{-7}\nonumber
\end{eqnarray}
\normalsize
One can hope that these conditions may be due to an internal structure of the 
mixing matrix $U_l$.
Therefore, we considered a CKM-like parametrization for the $U_l$ mixing matrix in terms of 
three angles $\theta_{13}^l$, $\theta_{12}^l$ y $\theta_{23}^l$,
\small
\begin{equation}\label{Ulparametrization}
 U_{l}=R(\theta_{23}^l)\cdot R(\theta_{13}^l)\cdot R(\theta_{12}^l),
\end{equation}
\begin{equation}
R(\theta_{23}^l) =\left(\begin{array}{ccc}
 1&0&0\\
 0&\cos\theta^{l}_{23}&\sin\theta^{l}_{23}\\
0&-\sin\theta^{l}_{23}&\cos\theta^{l}_{23}
\end{array}\right) ,
\end{equation}
\begin{equation}
 R(\theta_{13}^l)=\left(\begin{array}{ccc}
 \cos\theta^{l}_{13}&0&\sin\theta^{l}_{13}\\
 0&1&0\\
-\sin\theta^{l}_{13}&0&\cos\theta^{l}_{13}
\end{array}\right),
\end{equation}
\begin{equation}
R(\theta_{12}^l)=\left(\begin{array}{ccc}
 \cos\theta^{l}_{12}&\sin\theta^{l}_{12}&0\\
 -\sin\theta^{l}_{12}&\cos\theta^{l}_{12}&0\\
0&0&1
\end{array}\right).
\end{equation}
\normalsize
Then, just naming $U'$ instead of $U_{PMNS}$, we find the following relations
\small
\begin{equation}\label{equ1}
 \sin^{2}\theta_{13}=1-(U'_{11})^{2}-(U'_{12})^{2} \ ,
\end{equation}
\begin{equation}\label{equ2}
 \sin^{2}\theta_{23}=\frac{(U'_{23})^{2}}{(U'_{11})^{2}+(U'_{12})^{2}} \ ,
\end{equation}
\begin{equation}\label{equ3}
 \sin^{2}\theta_{12}=\frac{(U'_{12})^{2}}{(U'_{11})^{2}+(U'_{12})^{2}} \ .
\end{equation}
\normalsize
Evaluating the above equations in the ranges $\theta^l_{ij}\in [0,\pi/2]$, all the points $(\theta^l_{23},\theta^l_{13},\theta^l_{12})$ that matched 
the experimental allowed intervals are shown in Figure \ref{MatchMixAngles}.
We obtain a small region given by  two small angles and a large one: $ \theta^{l}_{12}=[0.06-0.15]$, $\theta^{l}_{13}=[0.07-0.16]$ and $\theta^{l}_{23}=[1.43-1.57\approx(\pi/2)]$. 
 Taking into account the allowed values for $(\theta^l_{23}$, $\theta^l_{13}$ y $\theta^l_{12})$, we can use a first-order 
approximation for the respective rotation matrices of $U_l$
\small
\begin{equation}\label{approx}
\approx\left(\begin{array}{ccc}
 1&0&0\\
 0&\epsilon&1\\
0&-1&\epsilon
\end{array}\right)\left(\begin{array}{ccc}
 1&0&\theta^{l}_{13}\\
 0&1&0\\
-\theta^{l}_{13}&0&1
\end{array}\right)\left(\begin{array}{ccc}
 1&\theta^{l}_{12}&0\\
 -\theta^{l}_{12}&1&0\\
0&0&1
\end{array}\right)
\end{equation}
\normalsize
where $\epsilon=\frac{\pi}{2}-\theta^{l}_{23} $.
Then, from equations (\ref{Msquare}) and (\ref{realMl2}) and  using the approximation (\ref{approx}),  we find the following model independent relations:
\small
\begin{eqnarray}
 (M_l^2)_{11}&=&a^{2}+b^{2}+c^{2}\nonumber\\
             & \approx & m_{e}^{2} + m_{\mu}^{2} \theta_{12}^{l2} + m_{\tau}^2\theta_{13}^{l2}\\
 &&\nonumber\\
  (M_l^2)_{22}&=&d^{2}+e^{2}+f^{2}\nonumber\\
	      &\approx& m_{\tau}^2+m_{e}^2(-\epsilon\theta_{12}^l-\theta_{13}^l)^2+m_{\mu}^2(\epsilon-\theta_{12}^l\theta_{13}^l)^2\nonumber \\
              &\sim & m_{\tau}^2\\
 &&\nonumber\\
 (M_l^2)_{33}&=&g^{2}+h^{2}+i^{2}\nonumber\\
             &\approx& m_{\tau}^2 \epsilon^2+m_{e}^2(\theta_{12}^l-\epsilon\theta_{13}^l)^2 +m_{\mu}^2(-1-\epsilon\theta_{12}^l\theta_{13}^l)^2 \nonumber\\
             &\sim& m_{\mu}^2
\end{eqnarray}
\normalsize
These expressions explain the observations considered before and represent the main 
result of this analysis. 

\section{Conclusions}
Assuming that the neutrino mass matrix is diagonalized by the TBM matrix 
 we tested several zero textures for the charged lepton mass matrix looking for those which provided $U_{PMNS}$ values in accordance with data. From the analysis we identified 
 ten three-zero textures satisfying our assumptions and delimited the size range of their entries. One of these textures (the one with zeros in the diagonal) exhibited a different and interesting behavior restricting the mixing angle $\theta_{23}$ to be located in a small range to agree with data. 
 Proposing a CKM-like parametrization of the lepton mixing matrix, $U_l$, we obtained some texture-independent relations in terms of the three rotation angles in $U_l$ and the charged lepton masses. We restricted ourselves to the case of real charged lepton mass matrices and fixed the lepton CP violating phase, but a model independent analysis considering the lepton sector CP violating Dirac phase was later realized in \cite{Acosta:2014dqa}.

\section{Acknowledgments}
Work supported by CONACyT (Mexico) and the Spanish grants FPA2011-22975 and Multidark CSD2009-00064 (MINECO), and PROMETEOII/2014/084 (Generalitat Valenciana).


\nocite{*}
\bibliographystyle{elsarticle-num}

\end{document}